\documentclass[10pt,twocolumn,a4paper]{esaAI}

\title{Enhancing Solar Driver Forecasting with Multivariate Transformers}

\def\AuthorEmail{s.shurtado@alumnos.upm.es}

\author[1]{Sergio Sanchez-Hurtado\thanks{Corresponding author. E-Mail: \AuthorEmail}}
\author[1]{Victor Rodriguez-Fernandez}
\author[2]{Julia Briden}
\author[2]{Peng Mun Siew}
\author[2]{Richard Linares}

\affil[1]{Universidad Politécnica de Madrid, Madrid, Spain}
\affil[2]{Massachusetts Institute of Technology, Cambridge, USA}

\begin{document}


\makeCustomtitle

\begin{abstract}
In this work, we develop a comprehensive framework for F10.7, S10.7, M10.7, and Y10.7 solar driver forecasting with a time series Transformer (PatchTST). To ensure an equal representation of high and low levels of solar activity, we construct a custom loss function to weight samples based on the distance between the solar driver's historical distribution and the training set. The solar driver forecasting framework includes an 18-day lookback window and forecasts 6 days into the future. When benchmarked against the Space Environment Technologies (SET) dataset, our model consistently produces forecasts with a lower standard mean error in nearly all cases, with improved prediction accuracy during periods of high solar activity. All the code is available on Github \faicon{github} : \url{https://github.com/ARCLab-MIT/sw-driver-forecaster}.
\end{abstract}


\section{Introduction}
\let\thefootnote\relax\footnotetext{

\vspace{-2.8em}
\begin{center}
  \rule{1.5cm}{0.2pt}
\end{center}
\vspace{-1em}

Short-paper preprint admitted for oral presentation in SPAICE Conference 2024 on September 16th-19th. See \url{https://spaice.esa.int/}
}

Solar flares and coronal mass ejections (CMEs) drive space weather events in the Earth's atmosphere. When charged particles from flares or CMEs reach Earth, atmospheric heating and transient solar wind activity increase, sometimes resulting in geomagnetic and solar storms. With a history of such storms disrupting communications and power systems and significantly increasing atmospheric drag for Low Earth Orbit (LEO) satellites, accurate space weather forecasting presents a critical enabling technology for mitigating these outages and satellite conjunction risk \cite{emmert2015thermospheric}.

One of the largest storms of the Space Age occurred in mid-October to early November 2003, causing significant satellite communications losses and inhibiting Global Positioning System (GPS) operations \cite{emmert2015thermospheric}. More recently, 38 Starlink satellites were de-orbited in early February 2022 due to a geomagnetic storm \cite{baruah2024loss}. As the solar cycle maximum approaches, more geomagnetic storms, such as the very recent G5-class geomagnetic storm from May 2024, may occur \cite{wdc2024dst}.

To quantify levels of space weather activity, solar and geomagnetic drivers are derived to serve as inputs to atmospheric density models, including the empirical Jacchia-Bowman 2008 (JB2008) model \cite{bowman2008jb2008}. The solar indices F10.7, S10.7, M10.7, and Y10.7 map energy from solar irradiance sources to major thermospheric layers, serving as inputs to JB2008. F10.7 is a proxy for solar activity, represented by the solar radio flux at a wavelength of 10.7 cm and correlating with sunspot numbers and solar irradiance \cite{tobiska_solar_geomagnetic}. S10.7 indicates the integrated 26-34 nm solar irradiance, M10.7 represents the modified daily Mg II core-to-wing ratio, and Y10.7 is an extension of the previously developed XL10.7, which measures the daily energy that is deposited into the mesosphere and lower thermosphere, weighted with Lyman-$\alpha$, the major energy source during moderate and low solar activity \cite{tobiska_solar_geomagnetic}.

Predicting future geomagnetic and solar storms and evaluating their potential impacts requires accurate solar driver forecasts. To assess forecast performance for a given prediction framework, SET provides a benchmarking dataset using an archived data set spanning 6 years and 15,000 forecasts across Solar Cycle 24 \cite{benchmarkSET}. In this work, we employ a multivariate approach using a transformer deep neural network to learn the mapping from historical solar drivers to future drivers, benchmarking this architecture against the predicted solar indices provided by SET \cite{benchmarkSET}.


\subsection{Current literature comparison}
Our multivariate transformer-based approach to forecasting solar drivers significantly advances the field of space weather forecasting. Traditional methods often rely on statistical models and simpler machine learning techniques, which can struggle with the complex nature of solar data. Neural network models, such as those by Assaf et al. reviewed \cite{NN_review_sw_2023}, outperform conventional methods by capturing complex temporal dependencies and extracting meaningful features from high-dimensional data.

As a demonstration of these advancements, Stevenson et al. \cite{N-Beats_emma_2020} applied univariate deep learning architecture to space weather proxy forecasting, yielding substantial improvements over traditional methods. Similarly, Daniell and Mehta \cite{f10_NNE_forecast} employed neural network ensembles to enhance the accuracy of solar proxy forecasting, achieving improvements in prediction accuracy through the use of ensemble learning techniques and advanced data manipulation methods. Building on this foundation, Briden et al. \cite{transformer_atmospheric_density_2023} demonstrated the effectiveness of transformer-based architectures in atmospheric density forecasting. Their work underscored the capability of transformers to handle long-term dependencies and nonlinearities in atmospheric data, providing a robust and accurate framework.

Here we suggest a framework that integrates these state-of-the-art approaches by utilizing the PatchTST transformer model. Consequently, our framework represents a significant step forward in the field, aligning with and extending the capabilities demonstrated in current academic work \cite{probabilistic_NNE_forecast}, while also incorporating an open-source component.


\subsection{Multivariate models}

This work utilizes the PatchTST forecaster to predict a time horizon of solar drivers based on a pre-defined look-back \cite{patchTST}. By using patching and channel independence, the model preserves local semantic information while reducing complexity. Channel independence makes the model robust to inter-channel-dependent noise, and patching divides input into smaller sections, reducing memory usage and computational complexity of Transformer attention maps.

Given a time series history of look-back window length $L$ for the solar drivers $\mathbf{z}_{1:L}$, the transformer backbone $\mathbb{H}$ maps to the predicted solar driver state for horizon length $T$, $\hat{\mathbf{z}}_{L+1:L+T} = (\hat{z}_{L+1}, \dots, \hat{z}_{(L+T)})$. The number of patches for the input time series is defined as $N = \lfloor \frac{L-P}{S} \rfloor + 2$, where $P$ is the patch length and $S$ is the stride, or the non-overlapping region between two patches. When utilized, patching reduces memory usage and the computational complexity of the attention map quadratically by a factor of $S$ \cite{patchTST}.

The PatchTST architecture includes a vanilla Transformer encoder with a trainable linear projection \(W_p \in \mathbb{R}^{D \times P}\) that maps observations to a latent space of dimension \(D\) \cite{patchTST}. A learnable additive position encoding \(W_\text{pos} \in \mathbb{R}^{D \times N}\) tracks the temporal order of patches. The encoded patches, \(z_d\), are processed by the multi-head attention mechanism, where each head \(h = 1, \ldots, H\) transforms them into query matrices \(Q_h^{(i)} = (z_d^{(i)})^T W_{Q_h}\), key matrices \(K_h^{(i)} = (z_d^{(i)})^T W_{K_h}\), and value matrices \(V_h^{(i)} = (z_d^{(i)})^T W_{V_h}\). Here, \(W_{Q_h}, W_{K_h} \in \mathbb{R}^{D \times d_k}\) and \(W_{V_h} \in \mathbb{R}^{D \times D}\). The attention output \(O_h^{(i)} \in \mathbb{R}^{D \times N}\) is obtained by scaling the dot product of the query and key matrices by \(\sqrt{d_k}\) (where \(d_k\) is the dimension of the query and key vectors), applying the softmax function, and multiplying by the value matrices.



Multi-head attention blocks include BatchNorm1 layers and a feed-forward network with residual connections \cite{Loffe2015}. The final output \(z^{(i)} \in \mathbb{R}^{D \times N}\) is flattened with a linear head to predict space weather indices over the horizon time \(z^{(i)} = (\hat{z}^{(i)}_{L+1}, \dots, \hat{z}^{(i)}_{(L+T)}) \in \mathbb{R}^{1 \times T}\). Instance normalization mitigates distribution shifts between training and test data. For more details on the PatchTST architecture, see \cite{patchTST, tsai}.

\section{Results}
\subsection{Data Splitting and Loss Function}

As mentioned, we aim to develop a comprehensive framework for forecasting solar drivers. Here, we present the initial steps using the SET publication data for F10.7, S10.7, M10.7, and Y10.7 drivers, covering 1997 to the present \cite{solfsmydata}. We filled missing S10.7 values with a 14th-degree polynomial, similar to \cite{probabilistic_NNE_forecast}, which uses linear interpolation. This approach better captures the trend and non-linearities of the index. We selected the polynomial that minimized the Wasserstein distance with the true values, looking for the filled data to align closely with the actual distribution, preserving dataset's statistical properties. 

\begin{figure*}[!t]
	\centering
	\includegraphics[width=.95\textwidth]{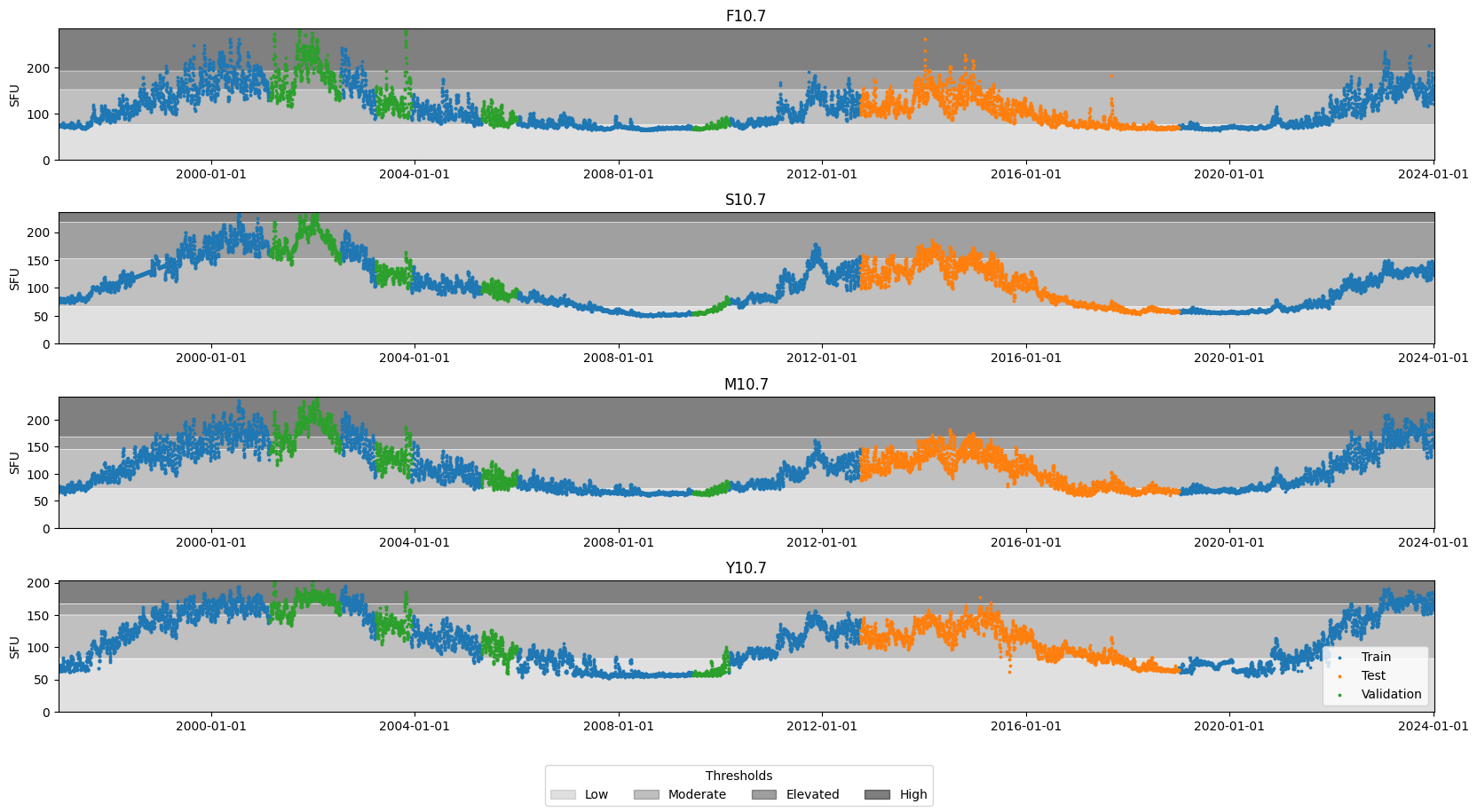}
    \caption{Distribution of F10.7, S10.7, M10.7, and Y10.7 data splits for training, validation and test sets, with solar activity thresholds shown in the background for each index.
    }
    \label{fig:data_splits}
\end{figure*}

In this work, we compare our results with the SET benchmark \cite{benchmarkSET}, which employs the SOLAR2000 statistical model. This model is strongly correlated with solar EUV irradiance and uses a linear predictive algorithm to identify persistence and recurrence patterns in solar activity. Future work will include benchmarks against other state-of-the-art forecasting methods to further assess model performance.

For comparison, we stablished the test-set as the period when SET performed their forecasts (2012 to 2018). The remaining dataset was divided into training and validation sets based on levels of solar activity. This novel data division strategy resulted from the challenge of establishing a validation set encompassing a full solar cycle and a training set with several cycles, which would be the optimal strategy for model training and validation. We first categorized the data into the four solar activity categories, as suggested by the authors of the benchmark \cite{benchmarkSET}. Then, we calculated the historical distribution of F10.7, the oldest solar driver, using data from 1947 to 1996 sourced from the National Oceanic and Atmospheric Administration (NOAA) \cite{f10historical}, while leaving out any data after 1997 to prevent data leakage. Then, we segmented the remaining dataset into fragments of equal size and searched for the best combination of them that most closely matched the historical distribution. Although this pattern is uncommon in time series analysis, temporal coherence was maintained, as these drivers model physical processes known to be cyclic. We experimented with various segment and validation set sizes until achieving the closest match. We aimed to make the validation split as small as possible to prevent significant loss of relevant patterns from the training set. This resulted in a segment size of 250 days, with 5 out of 30 segments designated for validation, leading to final split sizes of approximately 63\%/23\%/13\% (\cref{fig:data_splits}). In summary, by doing this we intend to have an accurate validation dataset that could confirm that the final model performs well in any situation.

\begin{figure}[!b]
    \centering
    \includegraphics[width=.95\columnwidth]{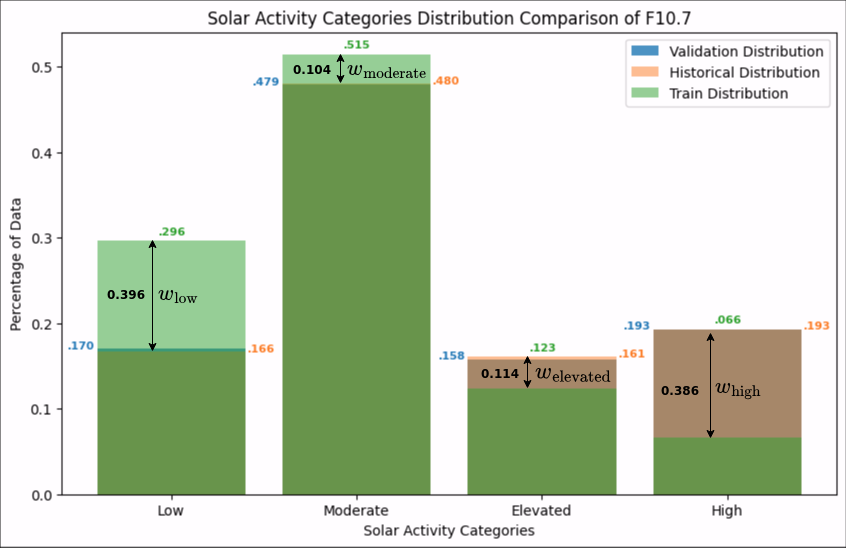}
	\caption{
         Comparison of solar activity for F10.7. \footnotesize{\textit{Note:} The weights ($w_s$  for $s \in S = \{\text{low, moderate, elevated, high}\}$) represent the differences between historical ($H_s$) and training ($T_s$) distributions, normalized to sum 1, defined as $w_s = |H_s - T_s| \cdot \frac{1}{\sum_{s \in S} |H_s - T_s|}$.}
    }
	\label{fig:distribution_comparison}
\end{figure}

\begin{equation}
\resizebox{0.9\columnwidth}{!}{$
\begin{aligned}
\text{wMSE} &= \frac{1}{N} \sum_{i=1}^{N} w_i (y_i - \hat{y}_i)^2 & \text{wMAE} &= \frac{1}{N} \sum_{i=1}^{N} w_i \left| y_i - \hat{y}_i \right|
\end{aligned}$}
\label{eq:MSE_MAE}
\end{equation}

\begin{figure*}[!h]
	\centering
	\includegraphics[width=.95\textwidth]{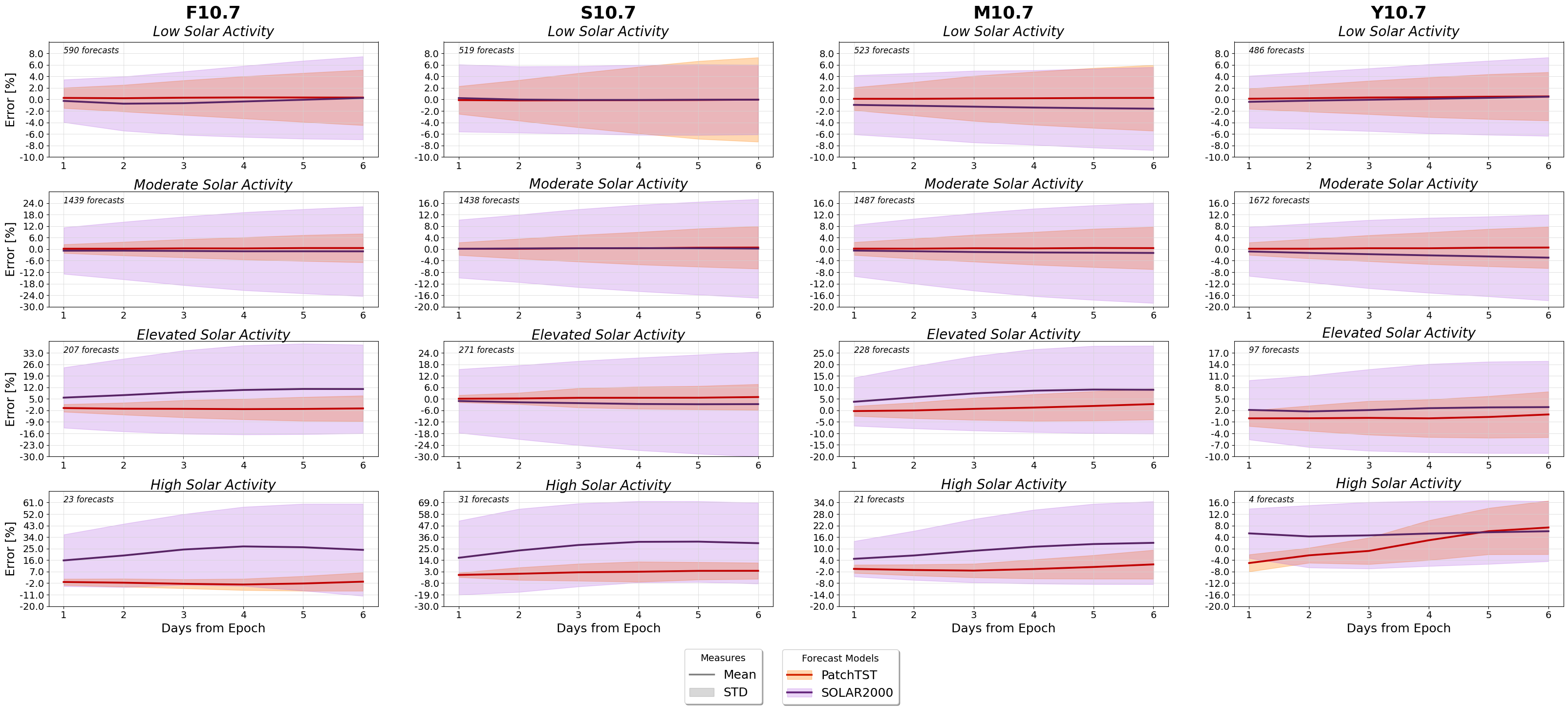}
    \caption{
        Comparison of SET benchmark against PatchTST ensemble with wMSE and wMAE losses.
        \footnotesize{\textit{Note:} The data is categorized with the SET dataset for better accuracy, specifically using different values for S10.7. However, we could not use the SET dataset for training as it only includes values between 2012 and 2018, which is insufficient for our training process.}
    }
    \label{fig:SET_Comparison}
\end{figure*}

With this splitting strategy, we under-represented and over-represented some of the solar activity categories, particularly high and low solar activity. To address this issue, we created a custom loss function for our model that is used when the distribution of samples is imbalanced between the training set and the "real" data distribution, as some authors have shown \cite{loss_func_clas, loss_func_analysis}. This custom loss function applies a weight to the loss calculation based on the distance between the pre-calculated F10.7 solar categories' historical distribution and the training set. The calculation of these weights can be seen in \cref{fig:distribution_comparison}. To implement this, we first categorize the target data, creating a tensor with weights for each category, matching the size of the original data. We then calculate the weighted error by multiplying this tensor with the squared or absolute error and apply a weighted average to obtain the weighted mean squared error (wMSE) and weighted mean absolute error (wMAE) [see \cref{eq:MSE_MAE}]. These measures are chosen to penalize large deviations in predictions, common during high solar activity, without excessively penalizing less volatile periods.

Sliding windows with a stride of 1, a horizon of 6 days (similar to the SET benchmark), and a look-back period of 18 days were used. This choice is based on the developers of PatchTST \cite{patchTST}, who claim that the architecture performs well with look-back sizes 2 to 3 times larger than the forecast horizon. For the model training, we require 30 epochs until the model learning stabilizes. We trained two models, one with wMAE as the loss function and another with wMSE. From these models, we created a simple ensemble by averaging the forecast values from both.

\subsection{Comparison with SET benchmark}

To show the initial performance of our approach, in this subsection, we will analyze its results against the benchmark data provided by SET \cite{benchmarkSET}. This comparison aims to highlight the strengths and potential areas for improvement. We utilize the same performance measures as SET: mean percentage error (MPE) and standard mean percent error (SMPE). By examining how our data aligns with benchmark (\cref{fig:SET_Comparison}), we observe that our model generally outperforms it, with a mean improvement of 77.72\% MPE and 60.22\% SMPE, resulting in significantly closer and accurate predictions. Notably, our model demonstrates significantly improved predictions during periods of high solar activity, traditionally one of the most challenging levels to forecast due to its volatility (83\% MPE and 66.5\% SMPE improvements). This aligns with findings from other researchers, who have noted that deep learning models tend to improve prediction accuracy for solar drivers as solar activity increases \cite{f10_high_accuracy, probabilistic_NNE_forecast, f10_NNE_forecast}.

When comparing the contribution of each loss, wMSE outperforms wMAE overall (88.93\% | 66.5\% MPE and 60.91\% | 59.53\% SMPE, respectively). However, we merge both as wMAE provides more stable improvements, especially at moderate levels (5.12\% MPE and 1.5\% SMPE better than wMSE). Meanwhile, wMSE significantly improves high solar activity forecast (8.55\% MPE | 1.1\% SMPE better than wMAE) and shows considerable improvement in low activity levels of F10.7 and S10.7 (53.1\% MPE | 2.5\% SMPE better on average).

However, it is important to note that our model does not perform as well at low solar activity levels. Although it still achieves better results on average than the benchmark, the difference is less pronounced. Additionally, it is noticeable that with a lower amount of data to be forecasted (such as in the case of Y10.7 during high  activity), our model's performance declines, as these approaches that benefit from large datasets.

\section{Discussion}
The initial results from our framework demonstrate promising performance compared to the SET benchmark. With the strategies mentioned, we achieve strong performance in forecasting the four solar drivers, especially during high solar activity periods. Traditional statistical models struggle in these periods due to volatility and insufficient information on active region growth, as noted by the benchmark authors \cite{benchmarkSET}. Beyond robustness, our model is efficient and lightweight, requiring less than two minutes of training (using two parallel dedicated Nvidia RTX3090 GPUs) to produce accurate forecasts. However, while our model excels during high solar activity periods, it shows comparatively less improvement during low levels, highlighting the need for further refinement.

We are exploring loss functions sensitive to low solar activity periods, including (i) classification loss functions that penalize prediction errors related to data volatility changes; (ii) mean squared logarithmic error (MSLE) and Huber loss functions for handling low-volatility data \cite{loss_func_survey}; and (iii) trended loss functions to provide contextual data trends. This last proposal also leads us to incorporate data trends as auxiliary model inputs, although it will involve preprocessing to integrate categorical and continuous data, enhancing model performance for cyclic data.

Additionally, we aim to nowcast S10.7, M10.7, and Y10.7 in our framework, using their current data to train a model based on F10.7 and extend the dataset back to 1947 for comprehensive training and validation. This approach has shown to be relevant in space weather forecasting \cite{nowcasting_analysis}, and some work has already been carried out using these techniques \cite{nowcast_EUV, radiation_belt_nowcast, images_nowcasting}.

Furthermore, we are integrating Ap and Dst indices into the PatchTST model, classifying their values per SET guidelines to apply custom weighted loss functions. Ap and Dst data, dating back to 1932 and 1957 respectively, will further enrich our work. Our framework will also include uncertainty measures and new test sets for comparison with recent works \cite{probabilistic_NNE_forecast}.

By benchmarking our PatchTST multivariate forecasting framework against the SET dataset for solar indices, we have demonstrated its superior performance, especially during high and moderate space weather activity. This advancement promises more reliable data, enhancing preparedness and response to space weather events. The improvements underscore its potential as a vital tool in mitigating the impacts of space weather on technology and infrastructure.
\printbibliography
\addcontentsline{toc}{section}{References}
\end{document}